\documentclass[a4paper,12pt]{amsart}

\usepackage{marginnote,geometry}
\usepackage{dsfont,amsmath,mathrsfs}
\usepackage[usenames,dvipsnames,svgnames,table]{xcolor}
\usepackage{cancel,amssymb}
\usepackage{graphicx}
\usepackage[pdfdisplaydoctitle=true,
            colorlinks=true,
            urlcolor=blue,
            citecolor=blue,
            linkcolor=blue,
            pdfstartview=FitH,
            pdfpagemode= UseNone,
            bookmarksnumbered=true]{hyperref}
\usepackage[color=blue!20, linecolor=red!]{todonotes}
\usepackage{lineno}
\newtheorem{theorem}{Theorem}[section]

\newtheorem{proposition}[theorem]{Proposition}

\theoremstyle{definition}

\theoremstyle{remark}

\numberwithin{equation}{section}
\newcommand{\ZZ}{\mathbb{Z}} % integer numbers
\newcommand{\RR}{\mathbb{R}} % real numbers
\newcommand{\id}{\mathds{1}}

\newcommand{\Ss}[1]{W^{k,p}(\mathbb{S}^1)}

\newcommand{\abs}[1]{\left\vert#1\right\vert}

\newcommand{\bra}[1]{\left\langle {#1}\right\vert}
\newcommand{\ket}[1]{\left\vert{#1}\right\rangle}

\DeclareMathOperator{\diag}{diag} %

\numberwithin{equation}{section}

\title[A dressing method for the Camassa-Holm equation]{A dressing method for soliton solutions of the Camassa-Holm equation}

\author{Rossen Ivanov}
\address{School of Mathematical Sciences, Dublin Institute of Technology, Kevin Street, Dublin 8, Ireland}
\email{rossen.ivanov@dit.ie}

\author{Tony Lyons}
\address{Department of Computing and Mathematics, Waterford Institute of Technology, Waterford, Ireland}
\email{tlyons@wit.ie}

\author{Nigel Orr}
\address{School of Mathematical Sciences, Dublin Institute of Technology, Kevin Street, Dublin 8, Ireland}
\email{nigel.orr@dit.ie }

\begin{document}
\begin{abstract} The soliton solutions of the Camassa-Holm equation are derived by the implementation of the dressing method. The form of the one and two soliton solutions coincides with the form obtained by other methods.
\end{abstract}

\maketitle

\section{Introduction}
In this paper we will develop a \textit{dressing method} to construct global solutions of the Camassa-Holm (or CH) equation, given by
\begin{equation}\label{eq:ch}
\left\{
\begin{aligned}
    q_t&+ 2u_x q + uq_x=0\\
    q&=u-u_{xx}
\end{aligned}
\right.\tag{CH}
\end{equation}
and where $\lim_{x\to \pm \infty }u(x,t)=u_0$, with $u_0$ a positive constant.
Equation (\ref{eq:ch}) is a bi-Hamiltonian system and admits interesting smooth and peaked traveling wave solutions \cite{CH93,CHH94}. It arises as a model equation in the study of two-dimensional water waves propagating over a flat bed \cite{CH93,CHH94,F95,DGH03,DGH04,J02,J03,J03a,CL09,HI11}. The Camassa-Holm equation has also been found to model the propagation of nonlinear waves in  cylindrical hyper-elastic rods, wherein $u(x,t)$ is interpreted as the radial stretching of a rod relative to the undisturbed state, see \cite{Dai98}.

As of now the volume of research papers dedicated to various aspects of the CH equation is probably measured in thousands and our references are by no means exhaustive. One remarkable feature of the Camassa-Holm equation (cf. \eqref{eq:ch}) is its peakon solutions, which are solutions of the form
\begin{equation}
    u(x,t) = q_0e^{-\abs{x-p_0t}}
\end{equation}
(where $\lim_{x\to \pm \infty }u(x,t)=0$) with $q_0$ and $p_0$ being constants, that is to say, weak solutions possessing a cusp at the wave crest, see \cite{CH93,CHH94,CS00,M2}. Physically the Camassa-Holm equation is of great interest since it allows for solutions displaying both peaking and breaking. The solutions with wave-breaking remain bounded but their gradient becomes unbounded in finite time, cf. \cite{CH93, CHH94, CE98, CE98a,C00,PS}. In addition to its versatility in modelling various physical phenomena, the Camassa-Holm equation embodies a rich mathematical structure, a particularly interesting feature in this regard being its interpretation as a geodesic flow on the Bott-Virasoro group \cite{M98,HMR98,CK03}.

The CH soliton solutions have been derived and studied by various methods, such as the Hirota method \cite{M1,M2,PI,PII,PIII}, the B\"acklund transform method \cite{RS,Li04}, the inverse scattering method \cite{CGI,BMS2}. In this study we extend the inverse scattering method by explicitly deriving the soliton solutions via the so-called dressing method.
The dressing method is one of the most convenient approaches to the derivation of the soliton solutions of integrable PDEs \cite{ZS1,ZS2,ZMNP,GVY}. The rationale of the method is the construction of a nontrivial (dressed) eigenfunction of the associated spectral problem from the known (bare) eigenfunction,
by means of the so-called dressing factor. The dressing factor is analytic in the entire complex plane, with the exception of the newly added simple pole singularities at pre-assigned discrete eigenvalues. The so called \textit{bare spectral problem}, is obtained by setting $u(x,t)\equiv u_0$=const and is trivially solved to provide for the bare eigenfunction. The \textit{dressing factor} will be the main object of our study.

The Camassa-Holm equation has many similarities with the integrable Degasperis-Procesi (or DP) equation \cite{DP,DHH}. The inverse scattering of the DP equation is studied in \cite{CIL,BMS1}, and in particular the dressing method for the DP equation is presented in \cite{CI16}.

 \section{The Spectral Problem for the Camassa-Holm Equation}\label{sec2}
\subsection{From the scalar to the matrix Lax pair}\label{sec2.1}
The Camassa-Holm equation can be represented as the compatibility condition $\phi_{xxt}\equiv\phi_{txx}$ for the solutions of the following spectral problem:
\begin{equation}\label{sec2.1eq1}
\left\{
\begin{aligned}
    &\phi_{xx}=\left(\frac{1}{4}+\lambda^2 q\right)\phi\\
    &\phi_{t}=\left(\frac{1}{2\lambda^2}-u\right)\phi_{x}+\frac{u_{x}}{2}\phi
\end{aligned}.
\right.
\end{equation}
In equation \eqref{sec2.1eq1} $\lambda$ is the spectral parameter, $\phi$ is a spectral eigenfunction while the potential $u(x,t)$ corresponds to a solution of the Camassa-Holm equation when $\lambda$ is time-independent. The compatibility produces also the relation $q=u-u_{xx}$. This solution $u$ may be obtained from the spectral problem above by means of the Inverse Scattering Transform as in \cite{CGI,CGI2}. The difference is that here we do not have a dispersion term $u_x,$ but instead we allow for a constant asymptotic value $u_0$ as $x\to \pm \infty$.

In our further considerations $u(x,\cdot)-u_0$ will be a Schwartz class function, where
$u_0 >0$ and with initial data $q(x,0) > 0$. Symmetry of the Camassa-holm equation then ensures that $q(x,t) > 0$ for all $t$ \cite{C01}. A discussion of the periodic case may be found in \cite{CM99} and \cite{C98}. Letting $k^{2}=-\frac{1}{4}-\lambda^2 u_0$, then the spectral parameter may be written as
\begin{equation} \label{lambda} 
\lambda^2(k)= -\frac{1}{u_0}\left(k^{2}+\frac{1}{4}\right),
\end{equation}
and the reader is referred to \cite{C01} for a discussion of the spectrum of the problem formed by equations \eqref{sec2.1eq1}--\eqref{lambda}. The continuous spectrum in terms of $k$
corresponds to $k\in\RR$. The discrete spectrum (in the upper half
plane) consists of finitely many points $k_{n}=i\kappa _{n}$,
$n=1,\ldots,N$ where $\kappa_{n}$ is real and $0<\kappa_{n}<1/2$, and hence, $\lambda_n=\lambda(i\kappa_n)$ is purely imaginary. Moreover there are two such eigenvalues, denoted by $\lambda_n=\pm i \omega_n$ where  $\omega_n>0.$

In the present work we will apply a variation of the Inverse Scattering Transform method, namely the dressing method, whereby a known solution is used to generate an new solution, thereby yielding a family of solutions of the Camassa-Holm equation. To implement this method it is first necessary to reformulate the spectral problem in \eqref{sec2.1eq1} as a \textit{matrix Lax pair}. To do so we define the eigenfunction $\phi_1$ as a solution of the spectral problem and we observe that the first member may be written according to
\begin{equation}\label{sec2.1eq2}
  \left(\partial-\frac{1}{2}\right)\left(\partial+\frac{1}{2}\right)\phi_{1}=\lambda^2 q\phi_{1}.
\end{equation}
This reformulation suggests the introduction of an auxiliary spectral function
\begin{equation} \label{sec2.1eq3}
    \phi_{2}:=\frac{1}{\lambda}\left(\partial+\frac{1}{2}\right)\phi_{1}
\end{equation}
from which it immediately follows that
\begin{equation*}
    \left(\partial-\frac{1}{2}\right)\phi_2 = \lambda q \phi_{1},
\end{equation*}
having applied equation \eqref{sec2.1eq2}. Defining the eigenvector
\[\Phi=\left(\begin{array}{cc}\phi_1\\ \phi_2\end{array}\right)\]
we may reformulate the spectral problem \eqref{sec2.1eq1} according to
\begin{equation}\label{sec2.1eq4}
\left\{
\begin{aligned}
&\Phi_{x}=\mathcal{L}\Phi\qquad \mathcal{L}:=\left(\begin{array}{cc}-\frac{1}{2} &\lambda \\ \lambda q & \frac{1}{2}\end{array}\right)\\
&\Phi_t=\mathcal{M}\Phi\qquad \mathcal{M}:=\left(\begin{array}{cc}\frac{1}{2}(u+u_x)-\frac{1}{4\lambda^2} &\frac{1}{2\lambda}-\lambda u\\\frac{1}{2\lambda}(q+u_x+u_{xx})-\lambda u q & \frac{1}{4\lambda^2}-\frac{1}{2}(u+u_x)\end{array}\right)
\end{aligned}
\right.
\end{equation}
which constitutes a matrix Lax pair for the Camassa-Holm equation. The compatability condition $\Phi_{tx}\equiv\Phi_{xt}$ for every eigenvector $\Phi$ immediatley implies the \textit{zero-curvature condition}, namely
\begin{equation}\label{sec2.1eq5}
    \mathcal{L}_{t}-\mathcal{M}_{x}+\left[\mathcal{L},\mathcal{M}\right]=0,
\end{equation}
where the bilinear operator $\left[\cdot,\cdot\right]$ denotes the usual matrix commutator. As with the scalar formulation of the spectral problem, comaprison of terms  of equal order in the spectral paramater $\lambda$ within the zero-curvature condition yields
\begin{description}
  \item[$\mathcal{O}(\lambda^{0})$] $u-u_{xx} =q$
  \item[$\mathcal{O}(\lambda^{1})$] $q_t+2u_x q+u q_x=0$,
\end{description}
which is precisely the Camassa-Holm equation.

%Since $q(x,t)>0$ for all $t$.

\subsection{The Gauge Transformed SL(2) Spectral Problem}\label{sec2.2}
We introduce the \text{gauge equivalent} matrix-valued eigenfunction $\Psi$ as follows
\begin{equation}\label{sec2.2eq1}
    \Phi=:G\Psi,
\end{equation}
where the gauge transformation $G$ is given by
\begin{equation}\label{sec2.2eq2}
    G = \left(\begin{array}{cc}
                q^{-\frac{1}{4}} & 0\\
                0 & q^{\frac{1}{4}}
                \end{array}\right).
\end{equation}
In terms of $\Psi,$ the spectral problem \eqref{sec2.1eq4} is written according to
\begin{equation}\label{sec2.2eq3}
\begin{aligned}
&\Psi_{x} = \tilde{L}\Psi\qquad \Psi_{t} =\tilde{M}\Psi,
\end{aligned}
\end{equation}
where we introduce
\begin{equation}\label{sec2.2eq4}
\tilde{L}:=G^{-1}\mathcal{L}G-G^{-1}G_{x}\qquad\tilde{M}:=G^{-1}\mathcal{M}G-G^{-1}G_{t}.
\end{equation}
In particular we find that the equation for $\Psi$ may be written as
\begin{equation}\label{sec2.2eq5}
\left\{
\begin{aligned}
&\Psi_x + (\tilde{h} \sigma_{3}-\lambda \sqrt{q} J) \Psi(x,t,\lambda)=0,\\
&\tilde{h} = \frac{1}{2}-\frac{q_x}{4q},\qquad J=\left(\begin{matrix}0&1\\1&0\end{matrix}\right)
\end{aligned}
\right.
\end{equation}
where $\sigma_3$ is the usual Pauli spin-matrix $\diag(1,-1)$. Changing from the $x-$variable to a new parameter defined by
\begin{equation}\label{y}
dy=\sqrt{q}dx, \qquad y=y(x,t),
\end{equation}
the spectral problem acquires the form of the standard Zakharov-Shabat spectral problem \cite{ZS1,ZS2,ZMNP,GVY}
\begin{equation}\label{ZS}
\begin{split}
&\Psi_y+ (h \sigma_3 -\lambda J) \Psi(y,t,\lambda) =0, \qquad h= \frac{1}{2\sqrt{q}}-\frac{q_y}{4q};\\
&L(\lambda)=h \sigma_3 -\lambda J \qquad \Leftrightarrow \qquad \Psi_y+ L(\lambda)\Psi=0.
\end{split}
\end{equation}
Since $L(\lambda)$ takes values in the algebra $sl(2)$, the eigenfunctions take values in the corresponding group - $SL(2).$ 

\subsection{Diagonlaisation}\label{sec2.3}
Upon imposing the trivial solution $u(x,t)\equiv u_0$ on the spectral problem we obtain the so-called \textit{bare} spectral problem, namely
\begin{equation}\label{sec2.3eq1}
\left\{
\begin{aligned}
&\Psi_{0,y}+(h_0\sigma_3-\lambda J)\Psi_0=0, \qquad h_0=\frac{1}{2\sqrt{u_0}},\\
&\Psi_{0,t}-\frac{1}{2h_0}\left(u_0-\frac{1}{2\lambda^2}\right ) (h_0\sigma_3-\lambda J)\Psi_0=0.
\end{aligned}
\right.
\end{equation}
Since $dy = \sqrt{u_0}dx$ then $y$ is simply a re-scaling of $x$ for the bare spectral problem. The solution of this linear system can be represented in the form
\begin{equation}\label{Psi0}
\Psi_0(y,t,\lambda)= V(\lambda) e^{-\sigma_3 \Omega(y,t,\lambda)}V^{T}(\lambda) C,
\end{equation}
where $C$ is an arbitrary constant matrix and
\begin{equation}\label{L-V}
\left\{
\begin{aligned}
&\Omega(y,t,\lambda)=\Lambda(\lambda) \left(y-\frac{1}{2h_0} \left(u_0 -\frac{1}{2\lambda^2}\right)t \right)\\
&\Lambda(\lambda)=\sqrt{h_0^2 + \lambda^2}\\
&V(\lambda)=\left(\begin{array}{cc}
        \cos \theta & -\sin \theta \\ \sin \theta & \cos \theta \\
\end{array}\right) \\
& \cos \theta =\sqrt{\frac{\Lambda+h_0}{2\Lambda}}, \qquad \sin \theta =\sqrt{\frac{\Lambda-h_0}{2\Lambda}}.
\end{aligned}
\right.
\end{equation}
In what follows $\Lambda$ will be always real and positive, however $\theta$ will be either real or imaginary.

\subsection{Symmetry reductions of the Spectral Problem}\label{sec2.4}
It can be verified easily that the spectral operator $L(\lambda)= h\sigma_3-\lambda J$ from \eqref{ZS}  possesses the following $\ZZ_2$-symmetry reduction (the bar is complex conjugation):
\begin{equation}\label{sec2.4eq1}
        \sigma_{3}\bar{L}(-\bar{\lambda})\sigma_3 = {L}(\lambda)
\end{equation}
since $ \sigma_3J\sigma_3=-J$. Likewise, the same relation is also true for the corresponding $M$-operator. This ensures that $h(y,t)$ is real. 
%\begin{equation}\label{sec2.4eq2}
%\sigma_{3}\widetilde{M}(-\lambda)\sigma_3 = \widetilde{M}(-\lambda).
%\end{equation}
Additionally, on the \textit{group valued} quantities, like the solutions $\Psi(y,t,\lambda)$ and the dressing factor $g$, (see the next section) we have 
\begin{equation}\label{Psisym1}
        \sigma_{3}\bar{\Psi}(y,t,-\bar{\lambda})\sigma_3 = {\Psi}(y,t,\lambda).
\end{equation}

%(to the spectral problem in equation \eqref{sec2.2eq3}--\eqref{sec2.2eq5} %we must also have
%\begin{equation}\label{sec2.4eq3}
%        \sigma_3\Psi(x,t,\lambda)\sigma_3 = \Psi(x,t,-\lambda).
%\end{equation}
%Following from the fact $\eta\in\RR$, we also deduce $\overline{\Psi}(x,\overline{\lambda}) %= {\Psi}(x,\lambda)$.

\noindent Analogously, noting that
\begin{equation}\label{sec2.4eq4}
J^2=\mathds{1},\qquad J\sigma_3J = -\sigma_3,
\end{equation}
%from which it follows at once that
%\begin{equation}\label{sec2.4eq5}
%J\widetilde{L}(\lambda)J = -\widetilde{L}(-\lambda).
%\end{equation}
%Meanwhile, we denote by $\widehat{\Psi}(x,t,\lambda)$ the group inverse %of the spectral function $\Psi(x,t,\lambda)$. 

\noindent and using $\Psi^{-1}(x,t,\lambda){\Psi}(x,t,\lambda)=\mathds{1}$, we have
\begin{equation}\label{sec2.4eq6}
\Psi^{-1}_y(\lambda) = \Psi^{-1}(\lambda)L(\lambda) \Rightarrow \Psi^{-1}_y(\lambda)^{T} = L(\lambda)\Psi^{-1}(\lambda)^{T},
\end{equation}
having also used $L^{T}(\lambda)=L(\lambda)$ in the last equation. Hence with \eqref{sec2.4eq4} we deduce
\begin{equation}\label{sec2.4eq7}
        \left(J\Psi^{-1}(\lambda)^{T}J\right)_y + L(-\lambda)\left(J\Psi^{-1}(\lambda)^{T}J\right)=0,
\end{equation}
that is to say $\Psi(-\lambda)$ and $J\Psi^{-1}(\lambda)^{T}J$ satisfy the same spectral problem, the solutions of which are unique (when fixed by the corresponding asymptotics in $y$ and $\lambda$), and thus $\Psi(\lambda) = J\Psi^{-1}(-\lambda)^{T}J$ or
\begin{equation}\label{sec2.4eq8}
\Psi^{-1}(y,t,\lambda) = J\Psi(y,t,-\lambda)^{T}J.
\end{equation} 

\section{The Soliton Solutions}\label{sec3}
\subsection{The Dressing Method}\label{sec3.1}
The $N$-soliton solution corresponds to a discrete spectrum containing $N$ distinct eigenvalues $\left\{\lambda_n: n=1,\ldots,N\right\}$.
The eigenfunctions of the spectral problem are singular at the discrete eigenvalues. Starting from a trivial ({\it bare}) solution $u(x,t)=u_0$ where $u_0$ is constant,
with corresponding eigenfunction $\Psi_0(x,t,\lambda)$, one may obtain an eigenfunction of a soliton solution $\Psi(x,t,\lambda)$ via the {\it dressing factor} $g:=g(x,t,\lambda)$,
\begin{equation}\label{sec3.1eq1}
        \Psi(x,t,\lambda) = g(x,t,\lambda) \Psi_0(x,t,\lambda).
\end{equation}
where $g$ is singular at the points of the discrete spectrum. We work with the $y$-representation as in equation \eqref{ZS}, where we have $x=X(y,t)$.

The dressing factor then satisfies the equation
\begin{equation}\label{sec3.1eq3}
        \partial_y g +h\sigma_3 g - gh_0 \sigma_3 -\lambda[J,g]=0 .
\end{equation}
Moreover, since the solution $\Psi(X(y,t),t,\lambda)\equiv \Psi(y,t,\lambda)$ belongs to the Lie group $SL(2)$, the  factor $g\in SL(2)$  and also satisfies the reductions given by equations  \eqref{Psisym1} and \eqref{sec2.4eq8}, namely
\begin{equation}\label{gsym}
        \sigma_{3}\bar{g}(y,t,-\bar{\lambda})\sigma_3 = {g}(y,t,\lambda), \qquad g^{-1}(y,t,\lambda) = Jg(y,t,-\lambda)^{T}J.
\end{equation}

We evaluate the spectral problem given by equation \eqref{sec2.2eq5}  at $\lambda=0$, and note that when written in terms of the $x$-variable has a solution
\begin{equation}\label{sec3.2eq1}
        \Psi(X(y,t),t,0)= e^{-\frac{1}{2}\left(X-\ln\sqrt{q}\right)\sigma_3}.
\end{equation}
In terms of the $y$-variable, the eigenfunction of the dressed spectral problem at $\lambda=0$ (cf. equation \eqref{ZS}), may be written as
\begin{equation}\label{sec3.2eq3}
\Psi(X(y,t),t,0) = g(y,t,0)\Psi_{0}(y,t,0)K_0,
\end{equation}
where $\Psi_{0}(y,t,0)$ is a solution of the bare spectral problem when $\lambda=0$, hence $K_0\in SL(2)$ is an arbitrary constant matrix, due to equation \eqref{sec3.1eq1}. 

We note however that when the $t$-dependence due to the second equation of \eqref{sec2.3eq1} is taken into account that $\Psi_{0}(y,t,\lambda)$ is singular at $\lambda=0$. As such, in equation \eqref{sec3.2eq3} we only consider the time-independent solution, namely, the solution which satisfies spectral problem associated with the $L$-operator. However, when $x\to \infty$ then $y\to \infty$ also, in which case
\[\Psi(X\to \infty,t,0)=\Psi_{0}(y=X\sqrt{u_0},t,0)K_0\]
and therefore $\Psi(X\to \infty,t,0) $ should be time-independent. Additionally, referring to equation \eqref{sec3.2eq1} we find \[\Psi_t(X,t,0)=-\frac{\sigma_3}{2}\left(X_t-\frac{q_t}{2q}\right)\Psi(X,t,0),\]
and  $\Psi_t(\infty,t,0)=-\frac{\sigma_3u_0}{2}\Psi(\infty,t,0),$ having imposed $X_t=u(X,t)\to u_0$ as $X\to\infty$. Hence $\Psi(X,t,0)$ must be of the form 
\[\Psi(X,t,0)=e^{-\frac{1}{2}\left(X-\ln\sqrt{q}-u_0 t\right )\sigma_3}\] 
in order to have the appropriate asymptotic behaviour (the correction being independent of  $y$). 
It follows that 
\begin{equation}\label{sec3.2eq2}
e^{-\frac{1}{2}\left(X(y,t)-\ln\sqrt{q}-u_0 t \right)\sigma_3}
= g(y,t,0) e^{-\frac{1}{2\sqrt{u_0}}\sigma_3y}K_0,
\end{equation}
which gives a differential equation for $X$ since $\partial_y X=q^{-1/2}$, cf. equation \eqref{y}. Thus it provides the change of the variables $x=X(y,t) $ in parametric form, where $y$ serves as the parameter. This of course is valid only in cases where the dressing factor is known and in what follows we shall explain how to construct it.

\subsection{Dressing factor with a simple pole}
In the $SL(2)$ Zakharov-Shabat spectral problems, the simplest form of $g$ possesses one simple pole \cite{ZMNP,GVY}, which leads to the following: 
\begin{proposition}\label{sec3.1prop1}
        The dressing factor $g(y,t,\lambda)$ is assumed to be of the form
        \begin{equation}\label{sec3.1eq4}
        g = \id+\frac{2i\omega A(y,t)}{\lambda-i\omega},\quad \text{where }\omega\in\RR
        \end{equation}
        and $A$ is a matrix-valued residue of rank 1.
\end{proposition}
By virtue of equation \eqref{gsym} and Proposition \ref{sec3.1prop1},  we deduce that the dressing factor must satisfy
\begin{equation}\label{sec3.1eq5}
\begin{aligned}
\left(\id + \frac{2i\omega A}{\lambda-i\omega}\right)\left(\id-\frac{2i\omega J A^TJ}{\lambda+i\omega}\right) =\id,
\end{aligned}
\end{equation}
and taking residues as $\lambda\to\pm i\omega$ we observe
\begin{equation}\label{sec3.1eq6}
\left\{
\begin{aligned}
&A\left(\id -JA^{T}J\right)=0\\
&\left(\id-A\right)JA^{T}J=0.
\end{aligned}
\right.
\end{equation}
Rewriting the matrix $A$ as
\begin{equation}\label{sec3.1eq7}
    A= \frac{\ket{n}\bra{m}}{\bra{m}n \rangle },\quad\text{with } \ket{n}=\left(\begin{matrix}n_1\\n_2\end{matrix}\right)\text{ and } \bra{m}=\left(\begin{matrix}m_1&m_2\end{matrix}\right),
\end{equation}
equations \eqref{sec3.1eq6}--\eqref{sec3.1eq7} combined with the symmetry relation \eqref{gsym} ensure
\begin{equation}\label{sec3.1eq8}
    \bra{n}=\frac{\bra{m}J}{\bra{m}J\ket{m}}\Rightarrow A=\frac{\ket{m}\bra{m}J}{\bra{m}J\ket{m}},
\end{equation}
in which case $A=A^2$ meaning $A$ is a projector.
Moreover, equation \eqref{sec3.1eq8} combined with the first symmetry relation of \eqref{gsym} also yields
\begin{equation}\label{sec3.1eq9}
 \sigma_3\bar{A}\sigma_3= A\Rightarrow m_1 = \mu,\ m_2=i\mu,\ \mu\in\RR.
\end{equation}
%\todotony{To Rossen: When I expand out the matrix relation I find
%$\frac{\overline{m}_2}{\overline{m}_1}=-\frac{m_2}{{m}_1}$
%which seems to suggest $m_1$ and $m_2$ only should have a phase difference of $\frac{\pi}{2}$. Is it by choice we let $m_1=\mu$ and $m_2=i\mu$?
%{\bf Yes Tony, it is our choice. Please do not use square and curly brackets. The square brackets are for the commutator}
%}

Replacing equation \eqref{sec3.1eq4} in equation \eqref{sec3.1eq3} and taking residues as $\lambda\to i\omega$ and $\lambda\to\infty$, we have
\begin{equation}\label{sec3.1eq10}
\begin{split}
(h&-h_0)\sigma_3=2i\omega [J,A], \\
  A_y& + \left(h_0\sigma_3+i\omega J\right)A-A\left(h_0\sigma_3-i\omega J\right)-2i\omega AJA=0.
\end{split}
\end{equation}
Replacing equation \eqref{sec3.1eq8} in equation \eqref{sec3.1eq10},
%we then find
%\begin{multline}\label{sec3.1eq11}
%\frac{\ket{m_y}\bra{m}}{\bra{m}J\ket{m}}+\frac{\ket{m}\bra{m_y}}{\bra{m}J\ket{m}}-\frac{2\bra{m}J\ket{m}\ket{m_x}\bra{m}}{\bra{m}J\ket{m}^2}+\frac{\left(h_0\sigma_3+i\omega J\right)\ket{m}\bra{m}}{\bra{m}J\ket{m}}\\
% -\frac{\ket{m}\bra{m}J\left(h_0\sigma_3-i\omega %J\right)J}{\bra{m}J\ket{m}}-\frac{2i\omega\bra{m}J^2\ket{m}\ket{m}\bra{m}}{\bra{m}J\ket{m}^2}=0
%\end{multline}
%having multiplied
multiplying everywhere by $J$ from the right and using $J\sigma_3J=-\sigma_3$ we have
\begin{multline}\label{sec3.1eq12}
\left(\ket{m_y}+\left(h_0\sigma_3+i\omega J\right)\ket{m}\right) \frac{\bra{m}}{\bra{m}J\ket{m}}+\frac{\ket{m}}{\bra{m}J\ket{m}}\left(\bra{m_y}+\bra{m}\left(h_0\sigma_3+i\omega J\right)\right)\\
-\frac{2\bra{m_y}J\ket{m}\ket{m}\bra{m}}{\bra{m}J\ket{m}^2}-\frac{2i\omega\bra{m}J^2\ket{m}\ket{m}\bra{m}}{\bra{m}J\ket{m}^2}=0.
\end{multline}
Assuming
\begin{equation}\label{sec3.1eq13}
 \bra{m_y}+\bra{m}\left(h_0\sigma_3+i\omega J\right) = 0
\end{equation}
we also observe that
\begin{equation}\label{sec3.1eq14}
-2\bra{m_y}J\ket{m}-2i\omega\bra{m}J^2\ket{m}=-2h_0\bra{m}\sigma_3J\ket{m}
\end{equation}
and using
\begin{equation}\label{sec3.1eq15}
\left.
 \begin{aligned}
  \bra{m}\sigma_3J\ket{m} = \bra{m}J\sigma_3\ket{m}\\
  J\sigma_3=-\sigma_3J
 \end{aligned}\right\}\Rightarrow \bra{m}\sigma_3J\ket{m} = 0,
\end{equation}
ensuring equation \eqref{sec3.1eq12} is satisfied identically provided \eqref{sec3.1eq13} holds. Furthermore, transposing equation \eqref{sec3.1eq13} we have
\begin{equation}\label{sec3.1eq17}
 \ket{m_y}+\left(h_0\sigma_3+i\omega J\right) \ket{m} = 0,
\end{equation}
an so $\ket{m}$ is an eigenvector of the {\it bare} spectral problem, in which case $\ket{m}$ is known.

\subsection{The one-soliton solution}
Equation \eqref{sec3.1eq17} suggest that $\ket{m}$ satisfies the bare spectral problem \eqref{sec2.3eq1} with $\lambda=-i\omega$, i.e. with spectral operator \[L_{0}(y,t,-i\omega)=h_0\sigma_3+i\omega J\]
Furthermore equation \eqref{sec3.1eq8} allows us to solve for $\ket{n}$ explicitly, thereby providing an explicit formula for the dressing factor $g(y,t,\lambda)$.
We can write the solution of \eqref{sec3.1eq17} as
\begin{equation}\label{sec3.2eq1}
  \ket{m}=\Psi(y,t,-i\omega)\ket{m_0}
\end{equation}
where $\ket{m_0}$ is a constant vector, and $\Psi(y,t,-i\omega)\in SL(2)$ satisfies the bare spectral problem
\begin{equation}\label{sec3.2eq2}
\left\{
\begin{aligned}
 &\Psi_{y} +L_{0}(y,t,-i\omega)\Psi=0\\
 &\Psi_{t} -v(\omega)L_{0}(y,t,-i\omega)\Psi=0\\
 &v(\omega)=\frac{1}{2h_0}\left(u_0+\frac{1}{2\omega^2}\right)
\end{aligned}
\right.
\end{equation}
With $\lambda=-i\omega,$  we have $\Lambda = \sqrt{h_0^2-\omega^2},$ 
\begin{equation}\label{sec3.2eq4}
 \sin(\theta)= i\sqrt{\frac{h_0-\Lambda}{2\Lambda}}\qquad \cos(\theta)= \sqrt{\frac{h_0+\Lambda}{2\Lambda}},
\end{equation} thus we conclude $\theta$ is imaginary. 

It follows from equation \eqref{L-V} that
\begin{equation}\label{sec3.2eq5}
\left\{
\begin{aligned}
 &\Psi(y,t,-i\omega)=V e^{-\sigma_3\Omega(y,t)}V^{-1}\\
 &\Omega(y,t)=\Omega(y,t,-i\omega)=\Lambda\left(y-\frac{1}{2h_0}\left(u_0+\frac{1}{2\omega^2}\right)t\right)
\end{aligned}
\right.
\end{equation}
while equation \eqref{sec3.2eq1} now ensures
\begin{equation}\label{sec3.2eq6}
 \ket{m}=\left(\begin{matrix}\mu_{01}e^{-\Omega(y,t)}\sqrt{\frac{h_0+\Lambda}{2\Lambda}}-i\mu_{02}e^{\Omega(y,t)}\sqrt{\frac{h_0-\Lambda}{2\Lambda}}\\ i\mu_{01}e^{-\Omega(y,t)}\sqrt{\frac{h_0-\Lambda}{2\Lambda}}+\mu_{02}e^{\Omega(y,t)}\sqrt{\frac{\eta_0+\Lambda}{2\Lambda}}\end{matrix}\right)
\end{equation}
where the coefficients $\mu_{01}$ and $\mu_{02}$ are defined as
\begin{equation}\label{sec3.2eq7}
  V^{-1}\ket{m_0}=\left(\begin{matrix}\mu_{01}\\ \mu_{02}\end{matrix}\right).
\end{equation}
Meanwhile, $\sigma_{3}\ket{m}=\ket{\bar{m}}$ yields $\mu_{01}=\bar{\mu}_{01}$ and $\mu_{02}=-\bar{\mu}_{02}$, while making the replacement $\nu_1=\frac{\mu_{01}}{\sqrt{2\Lambda}}$ and $i\nu_2=\frac{\mu_{02}}{\sqrt{2\Lambda}}$
we simplify $\ket{m}$ according to
\begin{equation}\label{sec3.2eq8}
  \ket{m}=\left(\begin{matrix}\nu_1e^{-\Omega(y,t)}\sqrt{h_0+\Lambda}+\nu_2e^{\Omega(y,t)}\sqrt{h_0-\Lambda}\\ i\nu_1e^{-\Omega(y,t)}\sqrt{h_0-\Lambda}+i\nu_2e^{\Omega(y,t)}\sqrt{h_0+\Lambda}\end{matrix}\right).
\end{equation}
Referring to equations \eqref{sec3.2eq4} and \eqref{sec3.2eq8} we have 
\begin{equation}\label{sec4.1eq6}
\begin{split}
 \frac{dX}{dy}e^{X-u_0t}&=-e^{-\frac{y}{\sqrt{u_0}}}\left[\frac{(h_0+\Lambda)e^{\Omega(y,t)} +ce^{-\Omega(y,t)}}{(h_0-\Lambda) e^{\Omega(y,t)}+ce^{-\Omega(y,t)}}\right]^2,\qquad c:=\frac{\nu_2\omega}{\nu_1} 
\end{split}
\end{equation}
haveing let $K_0= - \id$. 

Implementing the change of variables $(y,t)\to(-y,-t)$ and  observing that $\Omega(-y,-t)=-\Omega(y,t)$,  the differential equation for $X$ becomes
\begin{equation}\label{sec4.1eq6a}
\begin{split}
\frac{dX}{dy}e^{X-u_0t -\frac{y}{\sqrt{u_0}}}&=\left[\frac{(h_0+\Lambda)e^{-\Omega(y,t)} +ce^{\Omega(y,t)}}{(h_0-\Lambda) e^{-\Omega(y,t)}+ce^{\Omega(y,t)}}\right]^2. 
\end{split}
\end{equation}
The reason for doing so is the following: The Camassa-Holm equation written in terms of the  $(y,t)$-variables (the so-called ACH equation-see for instance \cite{RS,PLA}) is invariant under $(y,t)\to(-y,-t)$. Thus choosing  any solution of the ACH equation and imposing the change of varibles $(y,t)\to(-y,-t)$, we obtain another solution once we determine $x=X(y,t)$. In other words, if $X(y,t)$ is a soltution of the Camassa-Holm equation then so to is $x=X(-y,-t)$. Moreover, with this change of vafiables we also have $x\to \frac{y}{\sqrt{u_0}}$ when $y\to \pm \infty$, cf. equation \eqref{y}. 

\begin{figure}[h!]
 \centering
 \includegraphics[width=0.5 \textwidth]{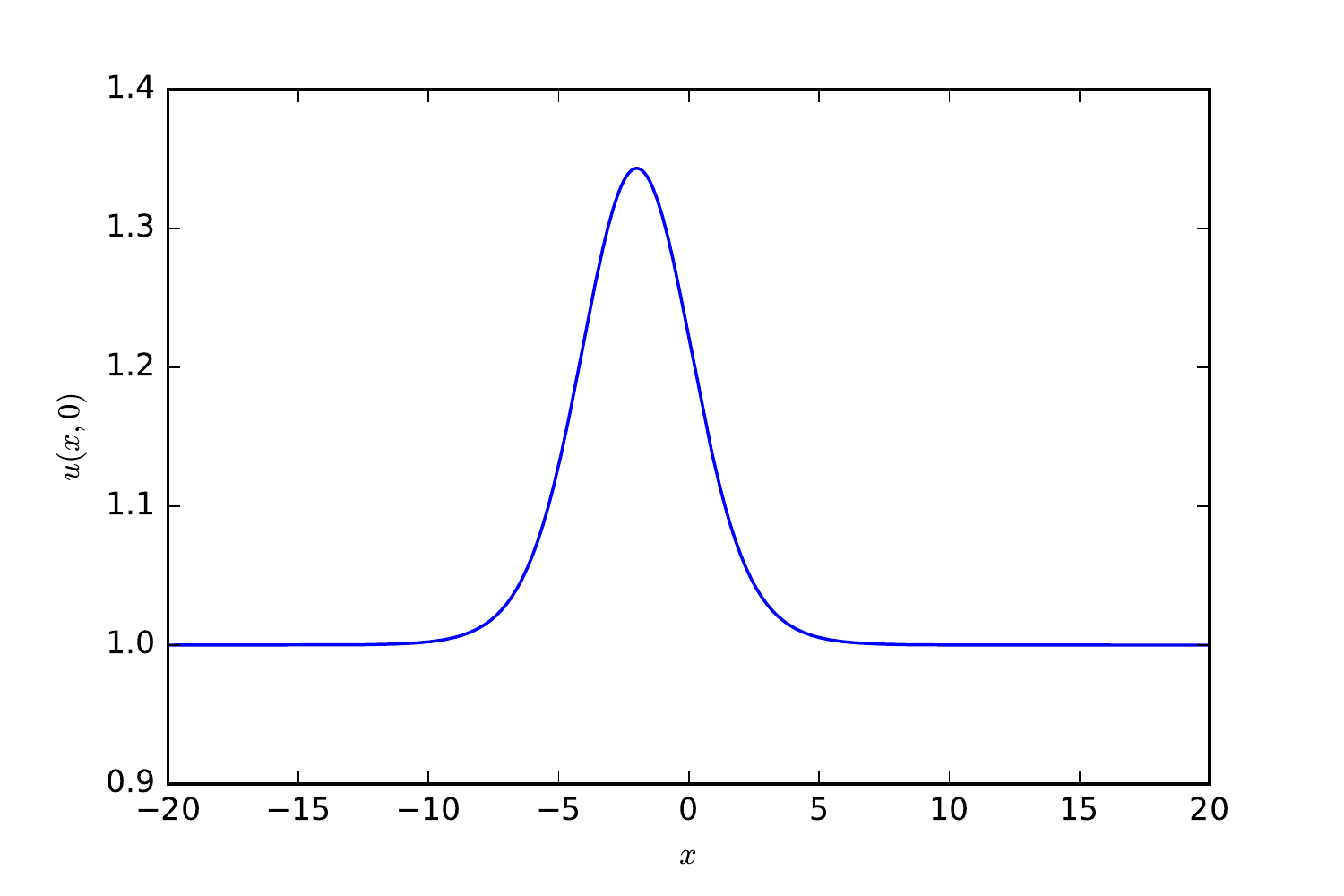}
 \caption{The one-soliton solution at $t=0$ where $\nu_1=1$, $\nu_2=2$, $u_0=1$ and $\omega=0.25$.}\label{fig1}
\end{figure}

Explicty the change of variables imposes the following trasnformation on our differential equation for $X$
\begin{equation}\label{yto-y}
\begin{split}
 \frac{dX(-y,-t)}{d(-y)}e^{X(-y,-t)-u_0(-t)}&=e^{-\frac{y}{\sqrt{u_0}}}\left[\frac{(h_0+\Lambda)e^{\Omega(y,t)} +ce^{-\Omega(y,t)}}{(h_0-\Lambda) e^{\Omega(y,t)}+ce^{-\Omega(y,t)}}\right]^2,\\
 \frac{dX(y,t)}{dy}e^{X(y,t)-u_0t}&=e^{\frac{y}{\sqrt{u_0}}}\left[\frac{(h_0+\Lambda)e^{\Omega(-y,-t)} +ce^{-\Omega(-y,-t)}}{(h_0-\Lambda) e^{\Omega(-y,-t)}+ce^{-\Omega(-y,-t)}}\right]^2,
\end{split}
\end{equation}
Formally this may be integrated by separation of variables, however we may also look for a solution in the form
\begin{equation}\label{sec4.1eq7}
 X(y,t)=\frac{y}{\sqrt{u_0}}+u_0t+\ln\left \vert\frac{A}{B}\right \vert
\end{equation}
with
\begin{equation}\label{sec4.1eq8}
 A=a_1e^{\Omega(y,t)}+a_2e^{-\Omega(y,t)}\qquad B=(h_0-\Lambda)e^{-\Omega(y,t)}+ce^{\Omega(y,t)}.
\end{equation}
Replacing equations \eqref{sec4.1eq7}--\eqref{sec4.1eq8} in equation \eqref{sec4.1eq6}, we conclude that $a_1=c$, $a_2=(h_0+\Lambda^{2})/(h_0-\Lambda)$ and thus
\begin{equation}
X(y,t)=\frac{y}{\sqrt{u_0}}+\ln\left\vert\frac{(h_0+\Lambda)^2e^{-\Omega(y,t)}
+\gamma^2e^{\Omega(y,t)}}{(h_0-\Lambda)^2e^{-\Omega(y,t)} +\gamma^2e^{\Omega(y,t)}}\right\vert , \qquad \gamma^2=c(h_0-\Lambda).
\end{equation}
When $\gamma^2=c(h_0-\Lambda)>0$ this expression can be written also as
\begin{equation}
X(y,t)=\frac{y}{\sqrt{u_0}}+u_0t+\ln\left\vert\frac{\frac{(h_0+\Lambda)^2-\gamma^2}{(h_0+\Lambda)^2+\gamma^2}\tanh {\Omega(y,t)}-1}{\frac{\gamma^2-(h_0-\Lambda)^2,}{\gamma^2+(h_0-\Lambda)^2} \tanh{\Omega(y,t)} +1}\right\vert.
\end{equation}
We introduc the constant
\[U=\frac{\Lambda}{h_0} <1\] 
and  choose constants such that $\gamma=\omega$, therby simplifying the expression for $X(y,t)$, which is now given by
\begin{equation}
X(y,t)=\frac{y}{\sqrt{u_0}}+u_0t+\ln\left\vert\frac{U\tanh {\Omega(y,t)}-1}
{U \tanh{\Omega(y,t)} +1}\right\vert
\end{equation}
The soliton solution itself is given in the form that appears in \cite{RS} as well
\begin{equation}
\begin{aligned}
u(X(y,t),t)&=\frac{\partial X}{\partial t};\\
           u(X,t)  &= u_0+ \frac{U^2\left(u_0+\frac{1}{2\omega^2}\right)(1-\tanh^2 \Omega)}{1-U^2 \tanh^2 \Omega},\\
\Omega&=\Lambda y-\frac{U}{2}\left(u_0+\frac{1}{2\omega^2}\right)t.
\end{aligned}
\end{equation}
We note that as $y\to\pm\infty$, then $\tanh \to \pm 1$ while $u\to u_0$, which we ovserve in the soliton profile shown in Figure \ref{fig1}.
Interestingly, choosing constants such that $c(h_0-\Lambda)=-\gamma^2<0$, $X(y,t)$ is no longer a monotonic function for all $y\in \mathbb{R}$, and the solution $u(x,t)$ is a function with discontinuities,  cf. \cite{RS} where such solutions are termed {\it unphysical}.

\subsection{The two-soliton solution}\label{sec4.2}

The dressing factor in this case has singularities at two different points of the discrete spectrum, which we denote $i\omega_1$ and $i\omega_2$, with residues $2i\omega_k A_k$ ($k=1,2.$). Extending Proposition \ref{sec3.1prop1}, we have
\begin{equation}\label{sec4.2eq1}
g(y,t,\lambda) = \id+\frac{2i\omega_1 A_1 (y,t)}{\lambda-i\omega_1}+\frac{2i\omega_2 A_2 (y,t)}{\lambda-i\omega_2}.
\end{equation}
The $\mathbb{Z}_2$ reduction $ \sigma_3 \bar{g}(y,t,-\bar{\lambda})\sigma_3= g(y,t,\lambda)  $ necessitates
\begin{equation}\label{sec4.2eq2}
\sigma_3 \bar{A}_k (y,t)\sigma_3= A_k(y,t), \qquad k=1,2.
\end{equation}
Applying equation \eqref{sec3.1eq3} to the dressing factor $g$ as given by equation \eqref{sec4.2eq1} ensures the corresponding equations for the residues, namely
\begin{equation}\label{sec4.2eq3}
A_{k,y}+h \sigma_3 A_k - A_k h_0\sigma_3 - i \omega_k [J, A_k]=0, \qquad k=1,2.
\end{equation}
Matrix solutions, $A_k$,  of the form
\begin{equation}\label{sec4.2eq4}
A_1 = \ket{n}\bra{m}, \qquad  A_2 = \ket{N}\bra{M},
\end{equation}
may be obtained if
\begin{equation}\label{sec4.2eq5}
\partial_y \ket{n} + (h\sigma_3-i\omega_1 J)\ket{n}=0, \qquad \partial_y\bra{m}= \bra{m} (h_0\sigma_3-i\omega_1 J)
\end{equation}
and
\begin{equation}\label{sec4.2eq5}
\partial_y \ket{N} + (h\sigma_3-i\omega_2 J)\ket{N}=0, \qquad \partial_y\bra{M}= \bra{M} (h_0\sigma_3-i\omega_2 J).
\end{equation}
That is to say, the vectors  $\bra{m},\bra{M}$ satisfy the bare equations and therefore are known in principle. The condition \eqref{sec4.2eq2} can be satisfied by imposing $\bra{\bar{m}}=\bra{m}\sigma_3$,  $\bra{\bar{M}}=\bra{M}\sigma_3$, and likewise for $\ket{n},\ket{N}$. The reduction given in \eqref{gsym} leads to
\begin{equation}\label{sec4.2eq6}
 \left[\id + \frac{2i\omega_1 A_1}{\lambda-i\omega_1}+\frac{2i\omega_2 A_2}{\lambda- i \omega_2}\right]\left[\id - \frac{2i\omega_1J A_1^{T}J}{\lambda+ i\omega_1}-\frac{2i\omega_2 J A_2^{T}J}{\lambda+i\omega_2}\right]=\id,
\end{equation}
which is identically satisfied for all $\lambda.$ Thus, the residues obtained at $\lambda=i\omega_k,$ (with $k=1,2$) ensure
\begin{eqnarray}\label{sec4.2eq7}
 A_1 \left(\id - J A_1^{T}J - \eta_2 J A_2^{T}J\right)& =&0,\qquad \eta_2=\frac{2\omega_2}{\omega_1+\omega_2},  \\  A_2 \left(\id -\eta_1 J A_1^{T}J -  J A_2^{T}J\right)& =&0,\qquad \eta_1=\frac{2\omega_1}{\omega_1+\omega_2}.
\end{eqnarray}
Using equations \eqref{sec4.2eq4}-\ref{sec4.2eq7} we obtain the following system
\begin{eqnarray}\label{sec4.2eq8}
 \bra{m}&=&\bra{m} J \ket{m} \bra{n} J+ \eta_2 \bra{m} J \ket{M} \bra{N} J , \\  \bra{M}&=&\eta_1 \bra{M} J \ket{m} \bra{n} J+  \bra{M} J \ket{M} \bra{N}J.
\end{eqnarray}
for the unknown vectors $\ket{n}$ and $\ket{N}$. The solutions are
\begin{eqnarray}\label{sec4.2eq9}
 \ket{n}&=&\frac{1}{\Delta}\big(\bra{M} J \ket{M} J\ket{m} - \eta_2 \bra{m} J \ket{M} J\ket{M} \big) , \\  \ket{N}&=&\frac{1}{\Delta}\big(\bra{m} J \ket{m} J\ket{M} - \eta_1 \bra{M} J \ket{m} J\ket{m} \big),
\end{eqnarray}
with
\begin{equation}\label{sec4.2eq10}
\Delta=\bra{M} J \ket{M}\bra{m} J \ket{m}-\eta_1 \eta_2\bra{m} J \ket{M}^2,
\end{equation}
which can be written in terms of the vector components as
\begin{equation}\label{sec4.2eq11}
\Delta=(\eta_1 m_2 M_1-\eta_2 m_1 M_2)(\eta_1 m_1 M_2 - \eta_2 m_2 M_1).
\end{equation}
Thus, the residues $A_k$ can be expressed in terms of the known vector components of \begin{equation}\label{mn} \bra{m}=\bra{m_{(0)}}\Psi_0^{-1}(y,t,i\omega_1),\qquad  \bra{M}=\bra{M_{(0)}}\Psi_0^{-1}(y,t,i\omega_2), \end{equation}
%\todo[inline]{Why $\Psi^{-1}_{0}$ here? $\Psi^{-1}_{0}$ satisfies the equation % $\partial_y \Psi^{-1}_{0} = \Psi^{-1}_{0} (h_0\sigma_3-\lambda J).$}
where $\bra{m_{(0)}},\bra{M_{(0)}} $ are arbitrary constant vectors.
The $SL(2)$ dressing factor \eqref{sec4.2eq1} at $\lambda=0$ is
\begin{eqnarray}\label{sec4.2eq12}
g(y,t;0)&=&\id - 2(A_1+A_2)=\mathrm{diag}(g_{11},g_{22})\\
&=&\mathrm{diag}\left(\frac{\omega_1 M_1 m_2 -\omega_2 M_2 m_1}{\omega_1 M_2 m_1-\omega_2 M_1 m_2},\frac{\omega_1 M_2 m_1-\omega_2 M_1 m_2}{\omega_1 M_1 m_2 -\omega_2 M_2 m_1}  \right), \nonumber
\end{eqnarray}
while the differential equation for $X(y,t)$ is
\begin{equation}\label{sec4.2eq13}
(\partial_y X) e^{X-2h_0 y-u_0 t} =g_{22}^2=\left(\frac{\omega_1 M_2 m_1-\omega_2 M_1 m_2}{\omega_1 M_1 m_2 -\omega_2 M_2 m_1}\right)^2.
\end{equation}
%We denote
%\begin{equation}
% \bra{m}=\bra{m_0}\Psi^{-1}(y,t,i\omega_1), \qquad \bra{M}=\bra{M_0}\Psi^{-1}(y,t,i\omega_2)
%\end{equation}
We recall that 
\begin{equation}
 \Psi(y,t,i\omega_{k})=V_{k}e^{-\sigma_3\Omega_k}V_k^{-1}
\end{equation}
with
\begin{equation}
\left\{
\begin{aligned}
 &\Omega_k(y,t)=\Lambda_{k}\left(y-\frac{t}{2h_0}\left(u_0+\frac{1}{2\omega_k^2}\right)\right),\quad\Lambda_k=\sqrt{h_0^2-\omega_k^2}\\
 &V_{k}=\left(\begin{matrix}\cos(\theta_k)&-\sin(\theta_k)\\ \sin(\theta_k)&\cos(\theta_k)\end{matrix}\right),\\
 &\cos(\theta_k)=\sqrt{\frac{h_0+\Lambda_k}{2\Lambda_k}}\qquad \sin(\theta_k)=i\sqrt{\frac{h_0-\Lambda_k}{2\Lambda_k}}
 \end{aligned}
 \right.
\end{equation}
for $k\in\{1,2\}$ (cf. equation \eqref{L-V}). We use \eqref{mn} noticing that $\bra{m_{(0)}}V_1= (\mu_1, i \mu_2)$ is a constant vector. Choosing $\mu_1, \mu_2$ to be real and positive, then explicitly  we have 
\begin{equation*}
\begin{split}
  m_1&=\sqrt{h_0+\Lambda_1}\sqrt{\frac{\mu_1}{\mu_2}}e^{\Omega_1(y,t)}+\sqrt{h_0-\Lambda_1}\sqrt{\frac{\mu_2}{\mu_1}}e^{-\Omega_1(y,t)},\\ m_2&=i\left(\sqrt{h_0-\Lambda_1}\sqrt{\frac{\mu_1}{\mu_2}}e^{\Omega_1(y,t)}+\sqrt{h_0+\Lambda_1}\sqrt{\frac{\mu_2}{\mu_1}}
e^{-\Omega_1(y,t)} \right),
 \end{split}
\end{equation*}
up to an irrelevant overall constant of $\sqrt{\mu_1 \mu_2}(2\Lambda_1)^{-1/2}$.

We may change the definition of $\Omega_1(y,t)$ by an additive constant,
\begin{equation}
    \Omega_1(y,t)=\Lambda_{1}\left(y-\frac{t}{2h_0}\left(u_0+\frac{1}{2\omega_1^2}\right)\right)+ \ln \sqrt{\frac{\mu_1}{\mu_2}},
\end{equation}
which yields
\begin{equation} \label{m12}
\begin{split}
  m_1&=\sqrt{h_0+\Lambda_1}e^{\Omega_1(y,t)}+\sqrt{h_0-\Lambda_1}e^{-\Omega_1(y,t)},\\ m_2&=i\left(\sqrt{h_0-\Lambda_1}e^{\Omega_1(y,t)}+\sqrt{h_0+\Lambda_1}e^{-\Omega_1(y,t)} \right)
 \end{split}
\end{equation}
Similarly, taking the constant vector $\bra{M_{(0)}}V_2= (\nu_1, -i \nu_2)$ with $\nu_1, \nu_2$ real and positive, we have
\begin{equation} \label{M12}
\begin{split}
  M_1&=\sqrt{h_0+\Lambda_2}e^{\Omega_2(y,t)}-\sqrt{h_0-\Lambda_2}e^{-\Omega_2(y,t)},\\ M_2&=i\left(\sqrt{h_0-\Lambda_2}e^{\Omega_2(y,t)}-\sqrt{h_0+\Lambda_2}e^{-\Omega_2(y,t)} \right),\\
 \Omega_2(y,t)&=\Lambda_{2}\left(y-\frac{t}{2h_0}\left(u_0+\frac{1}{2\omega_2^2}\right)\right)+ \ln \sqrt{\frac{\nu_1}{\nu_2}}.
 \end{split}
\end{equation}
The expression 
\begin{equation}
g_{22}=\frac{\omega_1 M_2 m_1-\omega_2 M_1 m_2}{\omega_1 M_1 m_2  -\omega_2 M_2 m_1}=\frac{T}{B},
\end{equation}
which we deduce from equations \eqref{m12}--\eqref{M12}, has denominator
\begin{equation}
\begin{split}
B=&\omega_1 \omega_2 \left(\frac{\Lambda_2 - \Lambda_1}{\sqrt{(h_0-\Lambda_1)(h_0-\Lambda_2)}}e^{\Omega_1+\Omega_2}+ + \frac{\Lambda_2 - \Lambda_1}{\sqrt{(h_0+\Lambda_1)(h_0+\Lambda_2)}}e^{-\Omega_1-\Omega_2}  \right. \\
  & + \left.  \frac{\Lambda_2 + \Lambda_1}{\sqrt{(h_0-\Lambda_1)(h_0+\Lambda_2)}}e^{\Omega_1-\Omega_2} + \frac{\Lambda_2 + \Lambda_1}{\sqrt{(h_0+\Lambda_1)(h_0-\Lambda_2)}}e^{-\Omega_1+\Omega_2} \right).
\end{split}
\end{equation}
Introducing the constants
\begin{equation} 
n_k=\sqrt[4]{\frac{h_0+\Lambda_k}{h_0-\Lambda_k}},
\end{equation} 
we obtain
\begin{equation}
B=\sqrt{\omega_1 \omega_2} (\Lambda_2^2-\Lambda_1^2) \left(n_1 n_2\frac{e^{\Omega_1+\Omega_2}}{\Lambda_1+\Lambda_2} +\frac{1}{n_1 n_2}\frac{e^{-\Omega_1-\Omega_2}}{\Lambda_1+\Lambda_2}
+\frac{n_1}{n_2}\frac{e^{\Omega_1-\Omega_2}}{\Lambda_2-\Lambda_1}+
\frac{n_2}{n_1}\frac{e^{-\Omega_1+\Omega_2}}{\Lambda_2-\Lambda_1} \right).
\end{equation}
Similarly it is found that
\begin{equation}
T=\sqrt{\omega_1 \omega_2} (\Lambda_2^2-\Lambda_1^2) \left(\frac{1}{n_1 n_2}\frac{e^{\Omega_1+\Omega_2}}{\Lambda_1+\Lambda_2} +n_1 n_2\frac{e^{-\Omega_1-\Omega_2}}{\Lambda_1+\Lambda_2}
+\frac{n_2}{n_1}\frac{e^{\Omega_1-\Omega_2}}{\Lambda_2-\Lambda_1}+
\frac{n_1}{n_2}\frac{e^{-\Omega_1+\Omega_2}}{\Lambda_2-\Lambda_1} \right).
\end{equation}
Again, we are looking for a solution of  \eqref{sec4.2eq13} in the form \eqref{sec4.1eq7}  with
\begin{equation}
A=\alpha_1 e^{\Omega_1+\Omega_2}+\alpha_2 e^{-\Omega_1-\Omega_2}+\alpha_3 e^{\Omega_1-\Omega_2}+\alpha_4 e^{-\Omega_1+\Omega_2},
\end{equation}
for some constants $\alpha_k,$ as yet unknown constants. Equation \eqref{sec4.2eq13}  requires
\begin{equation}
2h_0AB+BA_y-AB_y=T^2,
\end{equation}
whose solution $A$ is given by
\begin{equation}
A=\sqrt{\omega_1 \omega_2} (\Lambda_2^2-\Lambda_1^2) \left(\frac{1}{n_1^3 n_2^3}\frac{e^{\Omega_1+\Omega_2}}{\Lambda_1+\Lambda_2}+n_1^3 n_2^3\frac{e^{-\Omega_1-\Omega_2}}{\Lambda_1+\Lambda_2}
+\frac{n_2^3}{n_1^3}\frac{e^{\Omega_1-\Omega_2}}{\Lambda_2-\Lambda_1}+
\frac{n_1^3}{n_2^3}\frac{e^{-\Omega_1+\Omega_2}}{\Lambda_2-\Lambda_1} \right).
\end{equation}
The ratio $A/B$ may also be written as
\begin{equation}
\frac{A}{B}=\frac{1+ \frac{1}{n_1^6 }\frac{\Lambda_2+\Lambda_1}{\Lambda_2-\Lambda_1}e^{2\Omega_1}+ \frac{1}{n_2^6 }\frac{\Lambda_2+\Lambda_1}{\Lambda_2-\Lambda_1}e^{2\Omega_2} +\frac{e^{2\Omega_1+2\Omega_2}}{n_1^6n_2^6} }{1+ n_1^2\frac{\Lambda_2+\Lambda_1}{\Lambda_2-\Lambda_1}e^{2\Omega_1}+ n_2^2 \frac{\Lambda_2+\Lambda_1}{\Lambda_2-\Lambda_1}e^{2\Omega_2}+ n_1^2n_2^2e^{2\Omega_1+2\Omega_2} }
\end{equation}
which we simplyfy by means of the following redefinitions:
\begin{equation} \label{Omega12}
\begin{split}
 \Omega_1(y,t)&=\Lambda_{1}\left(y-\frac{t}{2h_0}\left(u_0+\frac{1}{2\omega_1^2}\right)\right)+ \ln \sqrt{\frac{\mu_1}{\mu_2}}-\ln n_1 + \frac{1}{2}\ln \frac{\Lambda_2+\Lambda_1}{\Lambda_2 - \Lambda_1},\\
 \Omega_2(y,t)&=\Lambda_{2}\left(y-\frac{t}{2h_0}\left(u_0+\frac{1}{2\omega_2^2}\right)\right)+ \ln \sqrt{\frac{\nu_1}{\nu_2}}-\ln n_2 + \frac{1}{2}\ln \frac{\Lambda_2+\Lambda_1}{\Lambda_2 - \Lambda_1}.
 \end{split}
\end{equation}
Alternatively, these may be simply written as
\begin{equation} \label{Omegak}
 \Omega_k(y,t)=\Lambda_{k}\left(y-\frac{t}{2h_0}\left(u_0+\frac{1}{2\omega_k^2}\right)\right)+\xi_k
\end{equation}
for some constants $\xi_k$ related to the initial separation of the solitons.  It follows that
\begin{equation}
\begin{split}
\frac{A}{B}&=\frac{1+ \frac{1}{n_1^4 }e^{2\Omega_1}+ \frac{1}{n_2^4 }e^{2\Omega_2} +\left(\frac{\Lambda_2-\Lambda_1}{\Lambda_2+\Lambda_1}\right)^2\frac{e^{2\Omega_1+2\Omega_2}}{n_1^4n_2^4} }{1+ n_1^4e^{2\Omega_1}+ n_2^4e^{2\Omega_2} + \left(\frac{\Lambda_2-\Lambda_1}{\Lambda_2+\Lambda_1}\right)^2n_1^4n_2^4e^{2\Omega_1+2\Omega_2} }\\
n_k^4&=\frac{1-U_k}{1+U_k}, \qquad U_k=\frac{\Lambda_k}{h_0}<1, \\
 X(y,t)&=\frac{y}{\sqrt{u_0}}+u_0t+\ln\left \vert\frac{A}{B}\right \vert , \\
u(X(y,t),t)&=\frac{\partial X}{\partial t}.
\end{split}
\end{equation}
which is of a form similar to that found in \cite{M1}. The two-soliton interaction is illustrated in Figure \ref{fig2} below.
\begin{figure}[h!] 
%\begin{tabular}{ccc}
\includegraphics[width=0.32\textwidth]{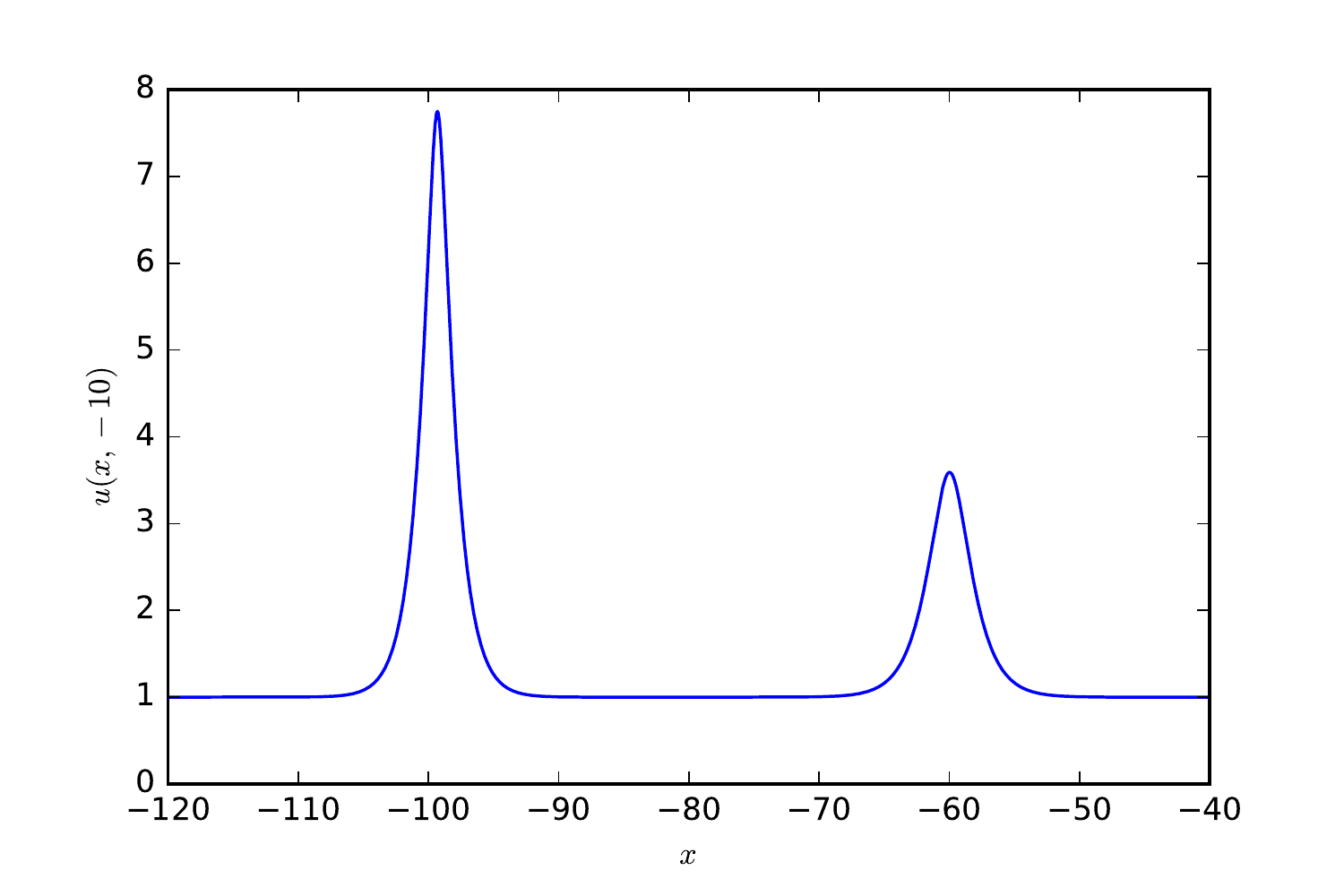}
\includegraphics[width=0.32\textwidth]{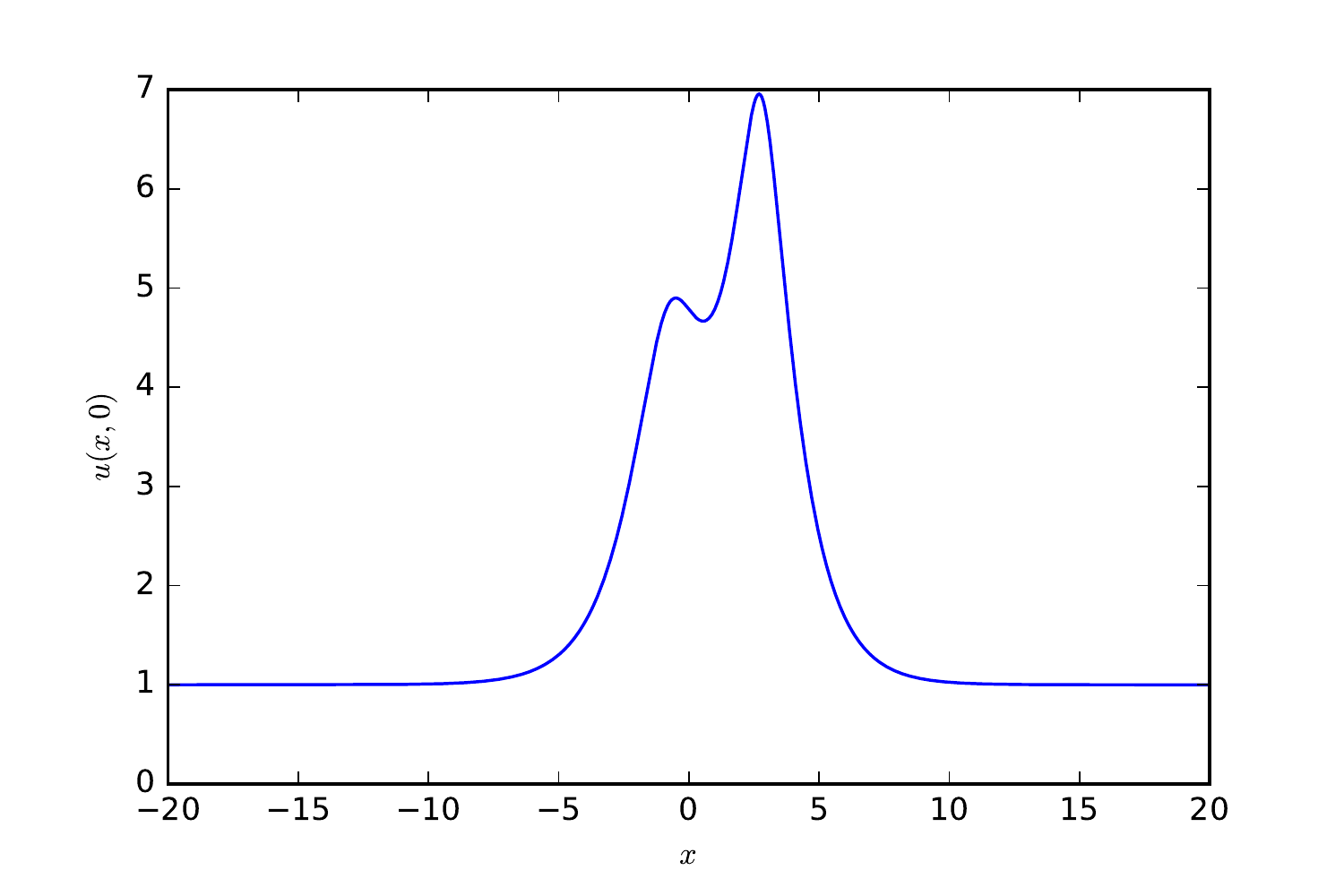}
\includegraphics[width=0.32\textwidth]{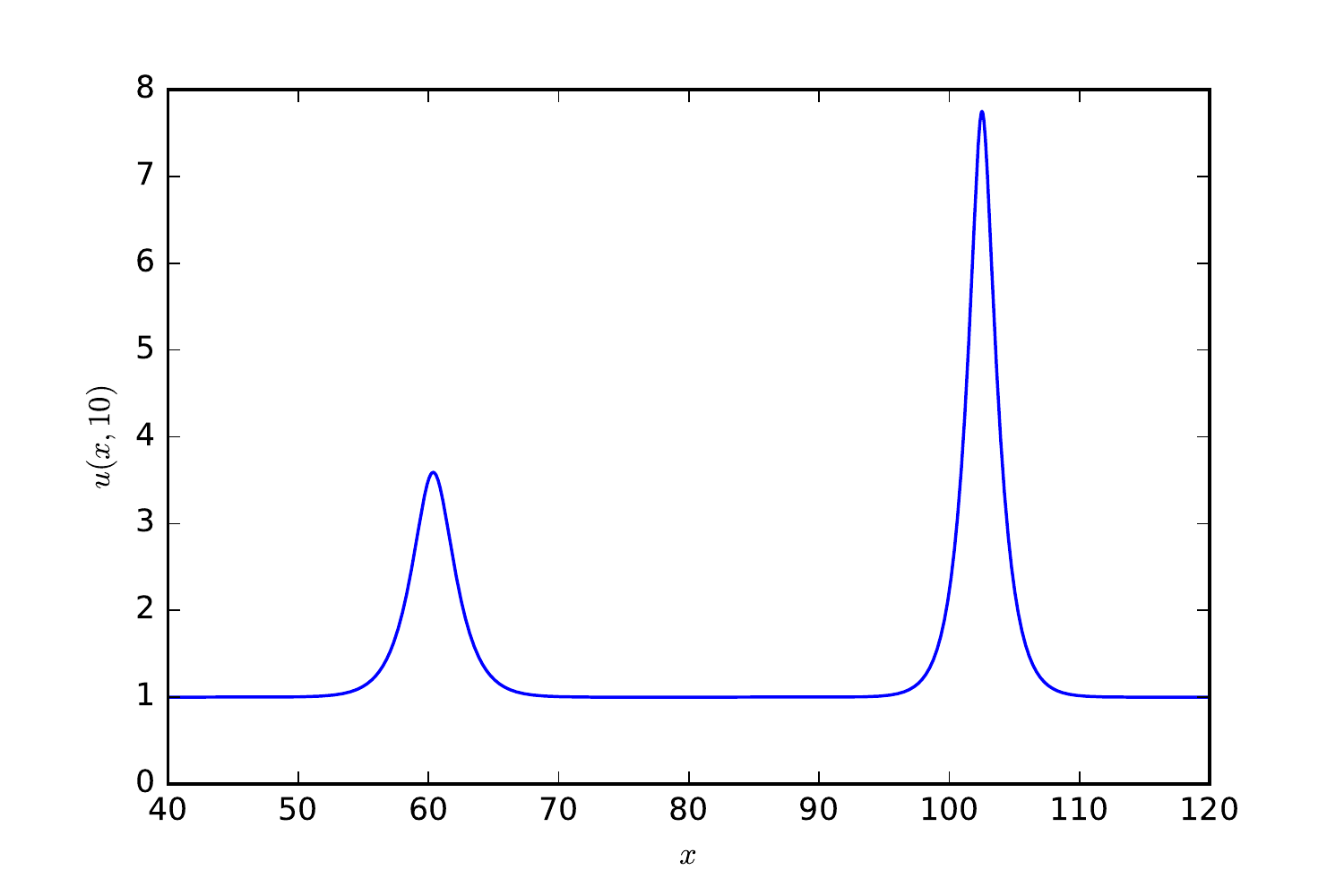}
%\end{tabular}
\caption{Snapshots of the two soliton solution of the Camassa-Holm equation (\ref{eq:ch}),
for three values of $t\in\{-10, 0, 10\}$. The other parameters are $u_0=1$, $\omega_1=0.35$ and $\omega_2=0.25$. The constants of integration were chosen as $\mu_1=1,\ \mu_2=2,\ \nu_1=2,\ \nu_2=3$.} \label{fig2}
\end{figure}

\section{Discussion}

We have applied the dressing method to derive the one and two soliton solutions of the Camassa-Holm equation. The multisoliton solutions can be obtained by other methods, however the dressing method is based on the spectral theory of the integrable system. The method can be extended for the multisoliton case by considering dressing factors with simple poles of the form   
\begin{equation}\label{secd.2eq1}
g(y,t,\lambda) = \id+\frac{2i\omega_1 A_1 (y,t)}{\lambda-i\omega_1}+\frac{2i\omega_2 A_2 (y,t)}{\lambda-i\omega_2}+\ldots+\frac{2i\omega_n A_n (y,t)}{\lambda-i\omega_n}.
\end{equation}

The restriction placed on the functional class (that is to Schwartz class) ensures the smoothness of the solutions for the corresponding discrete spectrum (and scattering data in general, which includes the choice of the constants $\mu_k$, $\nu_k$) . It is well known that in the limit $u_0 \to 0$ the solitons will develop a peak and become peakons \cite{RS}, see also \cite{M2,BMS2}.

We have to point out that for a different choice of the scattering data the dressing method provides the so-called cuspon solutions, which are characterised by waves with a cusp at the crest, where $u(x,t)$ is not differentiable. Such functions are solutions only in a week sense, and clearly outside of the Schwartz class. They will be obtained in a forthcoming publication.

%\subsection*{Acknowledgement}

\end{document}